\documentclass[a4paper,11pt]{article}
\pdfoutput=1 

\usepackage{jinstpub} 

\title{\boldmath Status of the TORCH Project}

\author[a,1]{N.~Harnew,\note{Corresponding author}}
\author[b,c]{S.~Bhasin,}
\author[e]{T.~Blake,}
\author[b]{N.H.~Brook,}
\author[f]{M.F.~Cicala,}
\author[g]{T.~Conneely,}
\author[c]{D.~Cussans,}
\author[d]{M.W.U.~van~Dijk,}
\author[d]{R.~Forty,}
\author[d]{C.~Frei,}
\author[e]{E. P. M.~Gabriel,}
\author[a]{R.~Gao,}
\author[f]{T.~Gershon,}
\author[d]{T.~Gys,}
\author[a]{T.~Hadavizadeh,}
\author[a]{T.~H.~Hancock,}
\author[f]{M. Kreps,}
\author[g]{J. Milnes,}
\author[d]{D.~Piedigrossi,}
\author[c]{J.~Rademacker}

\affiliation[a]{Denys Wilkinson Laboratory, University of Oxford, Keble Road, Oxford OX1 3RH, United Kingdom}
\affiliation[b]{University of Bath, Claverton Down, Bath BA2 7AY, United Kingdom}
\affiliation[c]{H.H. Wills Physics Laboratory, University of Bristol, Tyndall Avenue, Bristol BS8 1TL, United Kingdom}
\affiliation[d]{European Organisation for Nuclear Research (CERN), CH-1211 Geneva 23, Switzerland}
\affiliation[e]{School of Physics and Astronomy, University of Edinburgh, James Clerk Maxwell Building, Edinburgh EH9 3FD, United Kingdom}
\affiliation[f]{Department of Physics, University of Warwick, Coventry, CV4 7AL, United Kingdom}
\affiliation[g]{Photek Ltd., 26 Castleham Road, St Leonards on Sea, East Sussex, TN389 NS, United Kingdom}

\abstract{
The TORCH time-of-flight detector will provide particle identification between 
2-10\,GeV/$c$ momentum over a flight distance of 10\,m, and is 
designed for large-area coverage, up to 30\,m$^2$.   
A 15\,ps time-of-flight resolution per incident particle is anticipated by measuring the arrival times from  
Cherenkov photons  produced in a synthetic fused silica radiator plate of  10\,mm thickness. 
Customised Micro-Channel Plate Photomultiplier Tube (MCP-PMT) photon
detectors of 53 $\times$ 53\,mm$^2$ active area with a 64 $\times$ 64 granularity have been developed
with industrial partners. 
Test-beam studies using both a  small-scale TORCH demonstrator and a half-length TORCH module
are presented.  The desired timing resolution of
70\,ps per single photon is close to being achieved.  
}


\emailAdd{Neville.Harnew@physics.ox.ac.uk}

\proceeding{%
DIRC Workshop\\
11--13 Sept 2019\\
Castle Rauischholzhausen
}

\begin{document}
\maketitle
\flushbottom

\section{Introduction}
\label{sec:intro}

\begin{figure}
    \centering
    \includegraphics[width=0.28\linewidth]{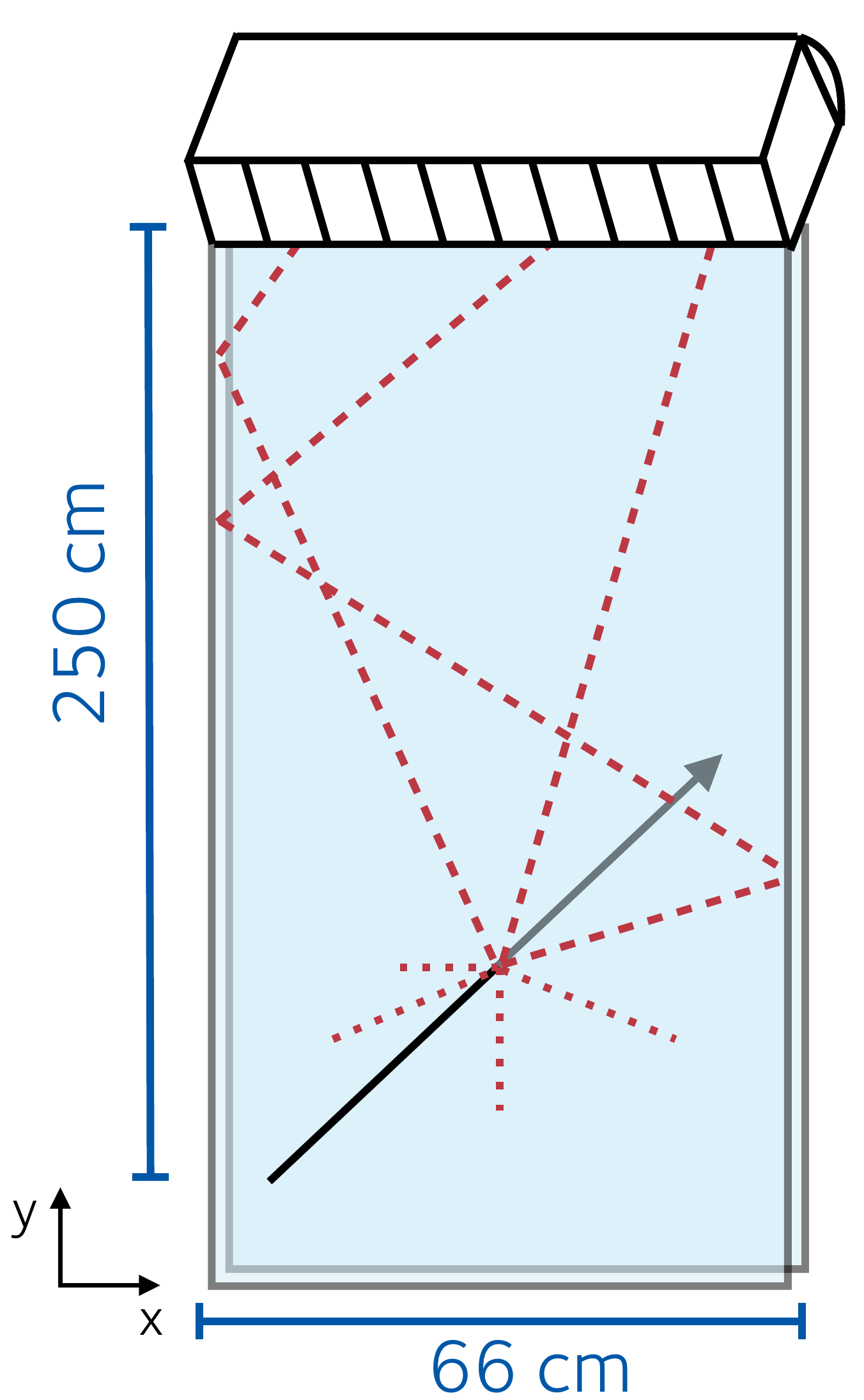}
    \includegraphics[width=0.32\linewidth]{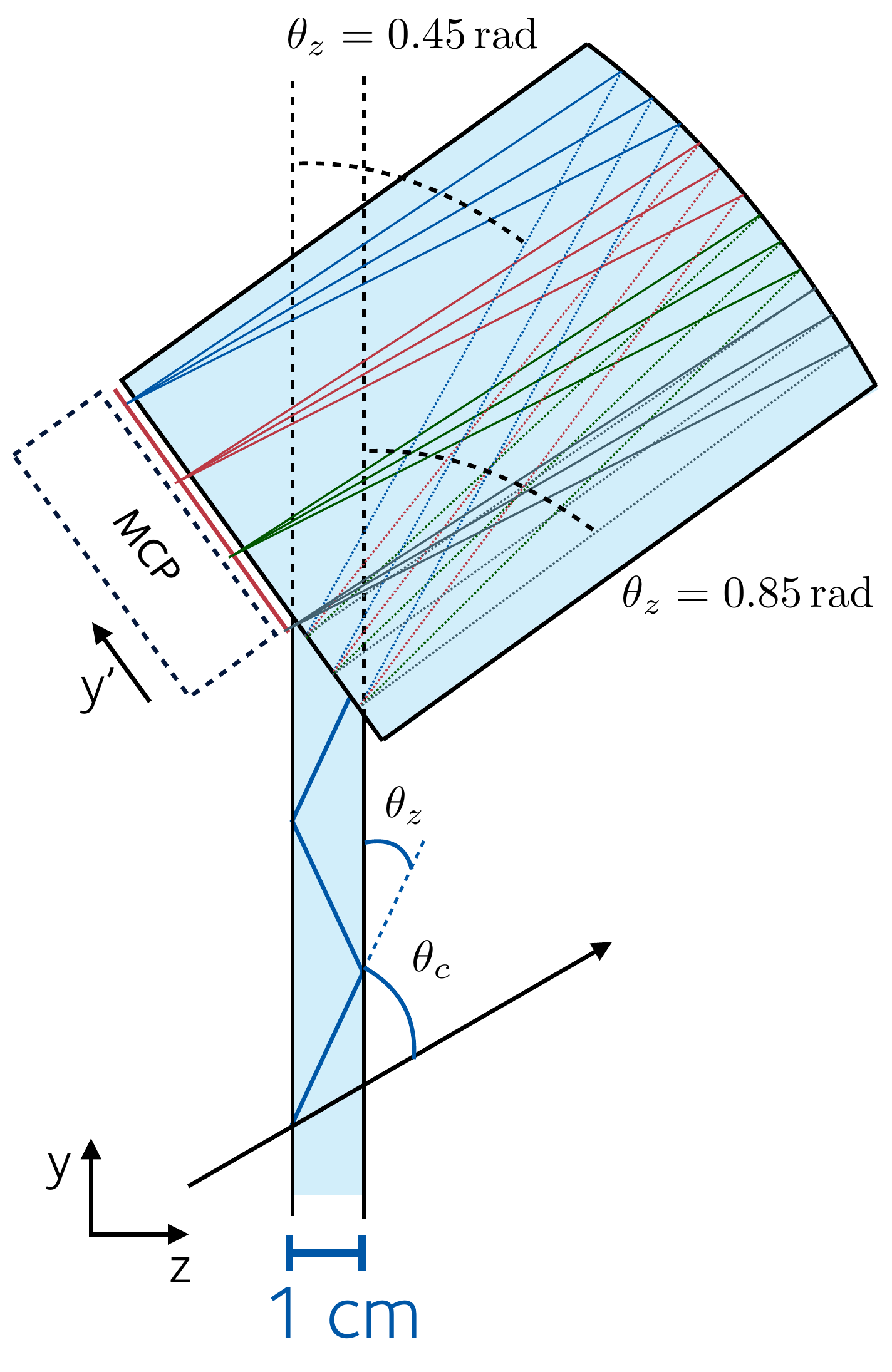}
    \includegraphics[width=0.33\linewidth]{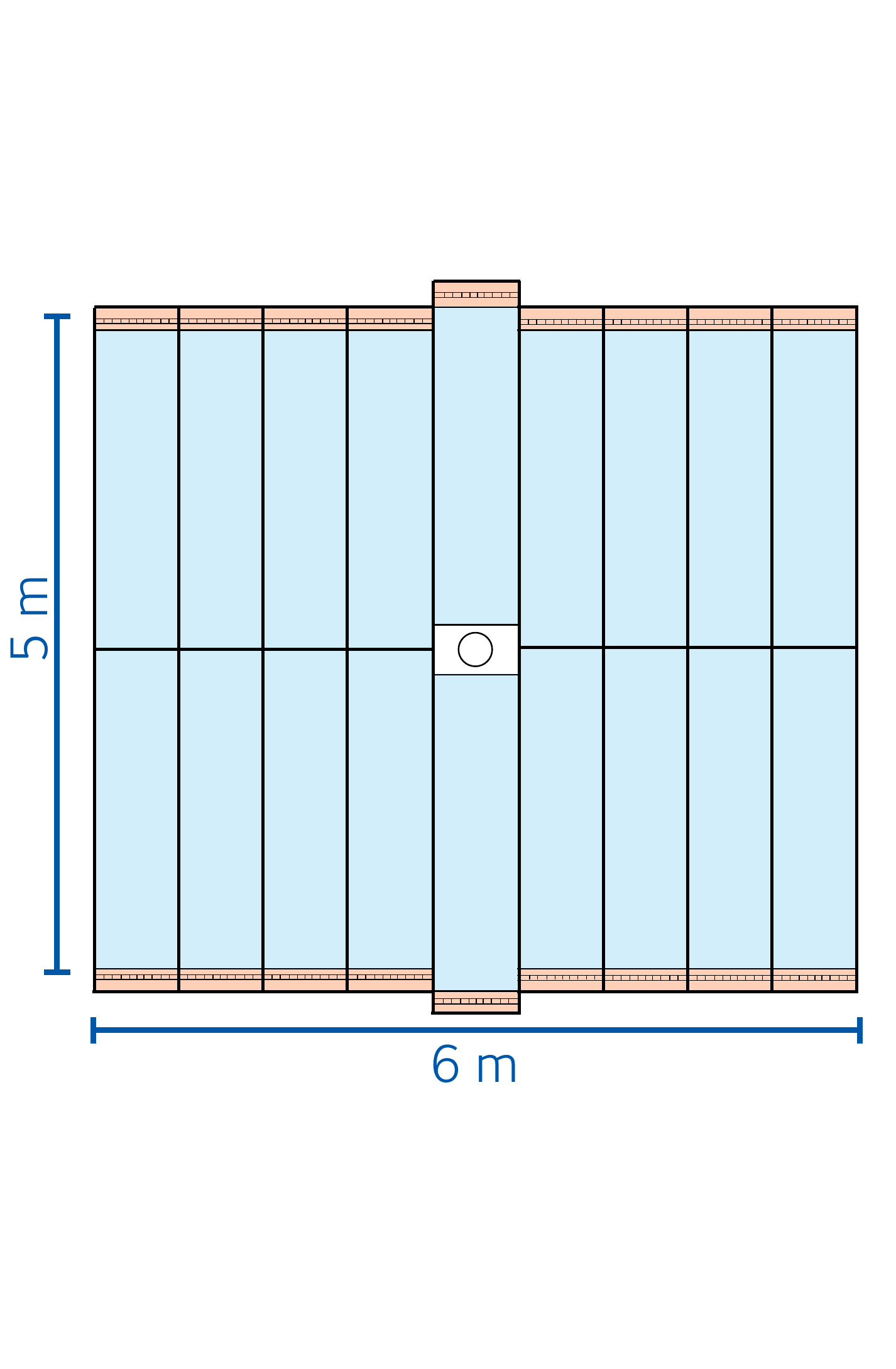}
    \caption{Schematics of the TORCH detector design. (Left) a single TORCH module, (middle) the focusing block, which translates the photon's angle of exit from the quartz plate in the \textit{yz} plane ($\theta_{z}$) into the \textit{y'} position on the MCP-PMT detector plane, and (right) the 18-module TORCH detector proposed for LHCb.}
    \label{fig:design_TORCH_module}
\end{figure}

The Timing Of internally Reflected CHerenkov photons (TORCH) is a high-precision time-of-flight detector suitable 
for large-area applications and covering a charged-particle momentum range up to 
10\,GeV/$c$\,\cite{ref:Charles2011, ref:Brook}. 
TORCH combines timing measurements with DIRC-style reconstruction, a technique pioneered by the  
BaBar DIRC\,\cite{ref:Babar} and  Belle\,II TOP\,\cite{ref:BelleII-TOP} collaborations.
Cherenkov photons, produced in a synthetic fused silica (quartz) radiator of 1\,cm thickness,
propagate by total internal reflection to focussing optics at the periphery of the detector.
There the photons are detected by fast photodetectors where their arrival times 
and propagation  angles are measured.  The Cherenkov angle is then reconstructed and chromatic dispersion is  accounted for in the determination of  the time of propagation of each photon\,\cite{ref:Brook}.
A schematic of the TORCH optics and modular layout is shown in Fig.\,\ref{fig:design_TORCH_module}. 

At 10\,GeV/$c$ momentum the time of flight difference between a pion and kaon is 35\,ps over a distance of 10\,m, 
hence a time resolution of about 10-15\,ps per track is required to give a 3-sigma separation. 
Given around 30 detected Cherenkov photons
per track, a time resolution of  $\sim$70\,ps per photon is therefore necessary. To achieve this, 
simulation has shown that a 1\,mrad  resolution of the angular measurement 
of each photon is required\,\cite{ref:Charles2011}. 

In this paper the operation of  both a small- and a large-scale  TORCH demonstrator module in 
a CERN test-beam will be described.

\section{The  photon detectors}
\label{sec:photodetectors}

The Cherenkov photons are focused onto Micro-Channel Plate Photomultiplier Tube (MCP-PMT) detectors.
The MCP-PMTs have been developed in collaboration with industrial partner Photek UK in a three-phase 
development programme\,\cite{ref:NHRICH2018}.
Each detector has a granularity of 64 $\times$ 64 pixels over a 53 $\times$ 53 mm$^2$ active area. A readout PCB is connected to the pixellated anode via Anisotropic Conductive Film.
Charge sharing and pixel grouping is  then employed to obtain an 
effective granularity of  8 $\times$ 128 pixels,  which is  required to achieve the 1\,mrad 
angular precision required by TORCH. 
The charge-sharing method relies on capacitive coupling to produce a smooth well-defined charge footprint on the anode\,\cite{ref:Conneely2}. 
This has the advantage of halving the number of anode pads in the confined area and also   the number of readout channels,  hence saving cost and complexity. 
In the coarse dimension, each of eight consecutive pixels are ganged together in the external electronics, giving  eight logical pixels. By similar grouping, tubes of effective granularity 4 $\times$ 128 pixels have also been employed.
The MCPs have ALD coating and are designed to withstand an integrated charge of 5\,C/cm$^2$\,\cite{ref:Conneely,
ref:Gys}.

\section{The TORCH  prototype modules and readout systems}
\label{sec:prototype}

Two prototype modules have been tested in a number of test-beam campaigns 
between 2015--2018 at the CERN PS T9 beamline.
The  first
is a small-scale TORCH demonstrator\,\cite{ref:Brook}, and consists of a 
$350 \times 120 \times 10$\, mm$^3$ (length, width and thickness) quartz radiator plate  with
a matching focusing block.
The demonstrator was instrumented with a single MCP-PMT,  with either a 4 $\times$ 64 or 8 $\times$ 64 
granularity.
%
%
%
The second demonstrator  is a  prototype of a half-length TORCH module, 
1250 $\times$ 660 $\times$ 10\,mm$^3$  (length, width and thickness). This module can be 
instrumented with ten MCP-PMTs, although only two 8 $\times$ 64 MCP-PMTs were equipped for the 
current test-beam running, corresponding to a total of 1024 readout channels.
For both demonstrators, the radiator plates were  glued to the
focussing blocks,  supported from  mounting frames, and enclosed in  light-tight vessels. 
The radiator plates were mounted in an almost vertical position, tilted backwards by 5$^\circ$ with respect to the horizontal  incidence of the beam. 

The two types of MCP-PMTs were read out by   independent fully-customised electronics 
readout systems\,\cite{ref:Gao2}. 
The systems comprise  front-end amplifier and discriminator boards, each of which  
houses two or four 32-channel NINO ASICs\,\cite{ref:Anghinolfi}.
Each NINO-32 board  is   connected to
one or two  TDC boards containing a pair of 
HPTDC chips\,\cite{ref:Christiansen} operated in 32-channel mode, which digitise the signals
with 100\,ps binning.
For the first small-scale demonstrator beam test with 4 $\times$ 64 MCP pixel readout, 
each NINO board had 64 channels instrumenting one column of pixels and which serviced a single HPTDC board.
 For the 8 $\times$ 64 MCP pixellization, 
the NINO boards were upgraded to have 128 channels instrumenting two 
columns of pixels and servicing a pair of HPTDC boards.

The NINO-32 provides time-over-threshold information which is used to correct 
time walk using a data-driven method\,\cite{ref:Brook}.
Non-linearities of the HPTDC time digitization  are also corrected.
Following these offline corrections, clustering due to the charge sharing  is  applied over 
neighbouring row MCP-PMT  hits, if the hits arrive within 1\,ns of each other, 
and the the centroid position of each photon hit then obtained.
However
the charge to width calibration has proven to be more challenging, and work is still ongoing to optimize this.


\section{Beam tests with the small-scale TORCH demonstrator}
\label{sec:small-scale}

The small-scale TORCH demonstrator was tested
in a 5 GeV/$c$ mixed pion/proton  beam in November 2017 and June 2018.
Proton-pion selection was achieved independent of TORCH using a pair of upstream Cherenkov counters.
Two borosilicate finger counters (T1 and T2) separated by  a $\sim$11\,m flight path provided a time reference.
The beam was incident approximately 
14\,cm below the plate centre-line and close to the plate side, 
a position which was chosen to give a cleanly resolved pattern.

\begin{figure}
\centering
\includegraphics[width=0.95\linewidth]{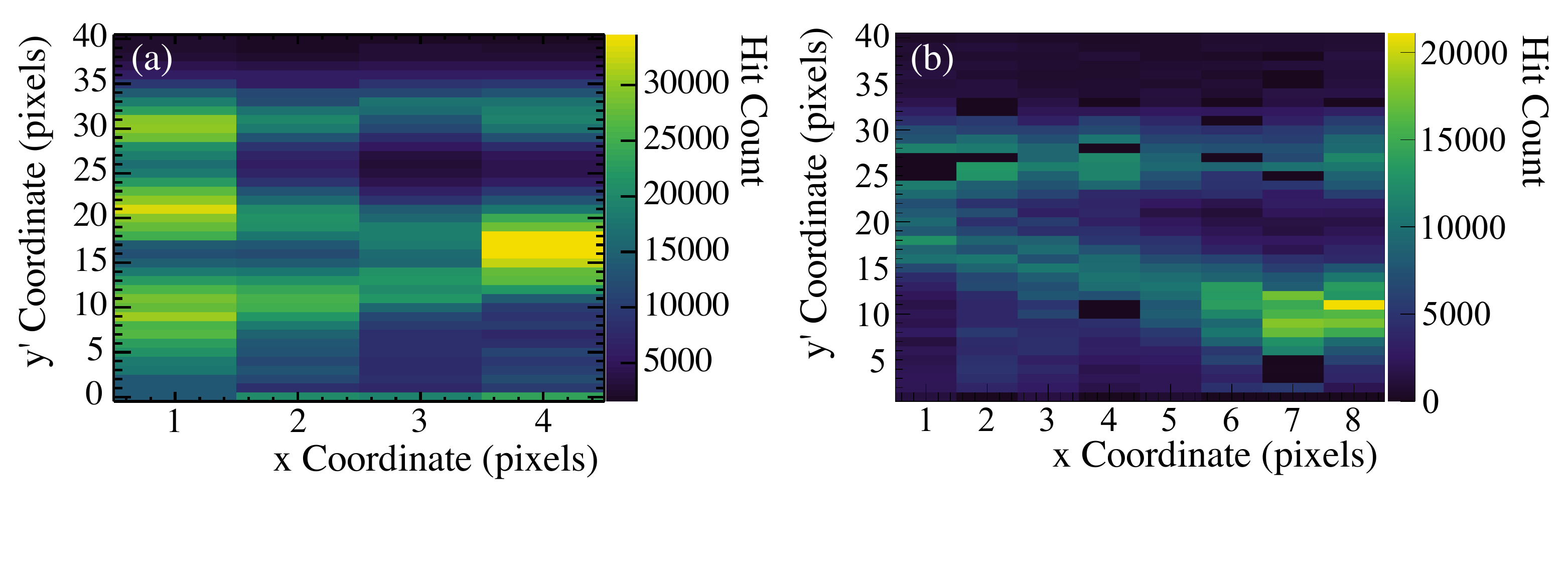}
\caption{The patterns of hits measured in  the TORCH small-scale demonstrator for a combined
 5\,GeV/$c$   pion and proton beam:
(a)   4 $\times$ 64      and
 (b)  8 $\times$ 64 pixel readout.}
\label{fig:hits}
\end{figure}

The patterns of measured  photon hits are shown for
both the  4 $\times$ 64  and   8 $\times$ 64 MCP-PMT configurations in   Fig.\ref{fig:hits}.
Hyperbola-like patterns at the MCP-PMT plane
result from the Cherenkov cones which are internally reflected in the quartz radiator; 
reflections off the module sides result in a folding of this pattern.
Chromatic dispersion spreads the lines into bands. Since only a single MCP-PMT is instrumented, the
full pattern is only sampled, which accounts for the  observed  discontinuities. 
The dead channels  observed in  the hit distribution of the 8 $\times$ 64 MCP-PMT  are due to NINO 
bonding issues (and which have since been rectified).

\begin{figure}
\centering
\hspace{0.5truecm} 
\includegraphics[width=0.75\linewidth]{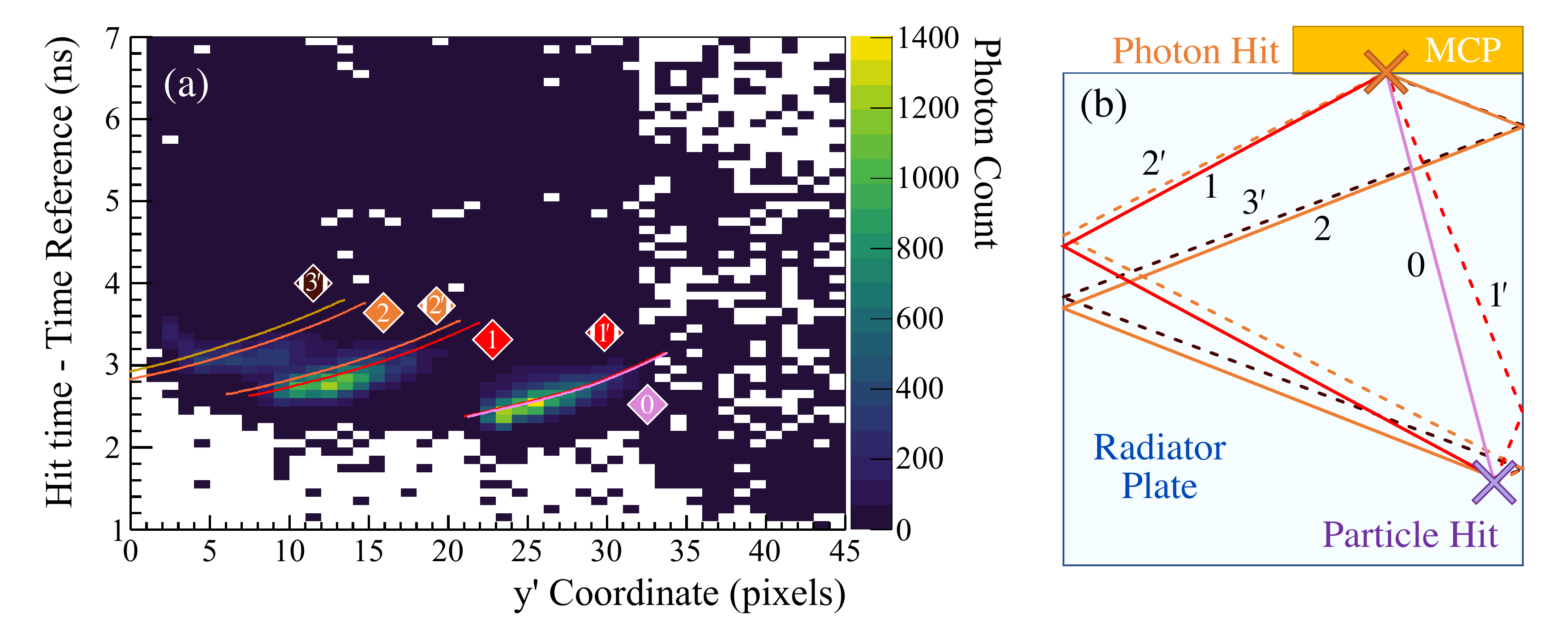}
\includegraphics[width=0.45\linewidth]{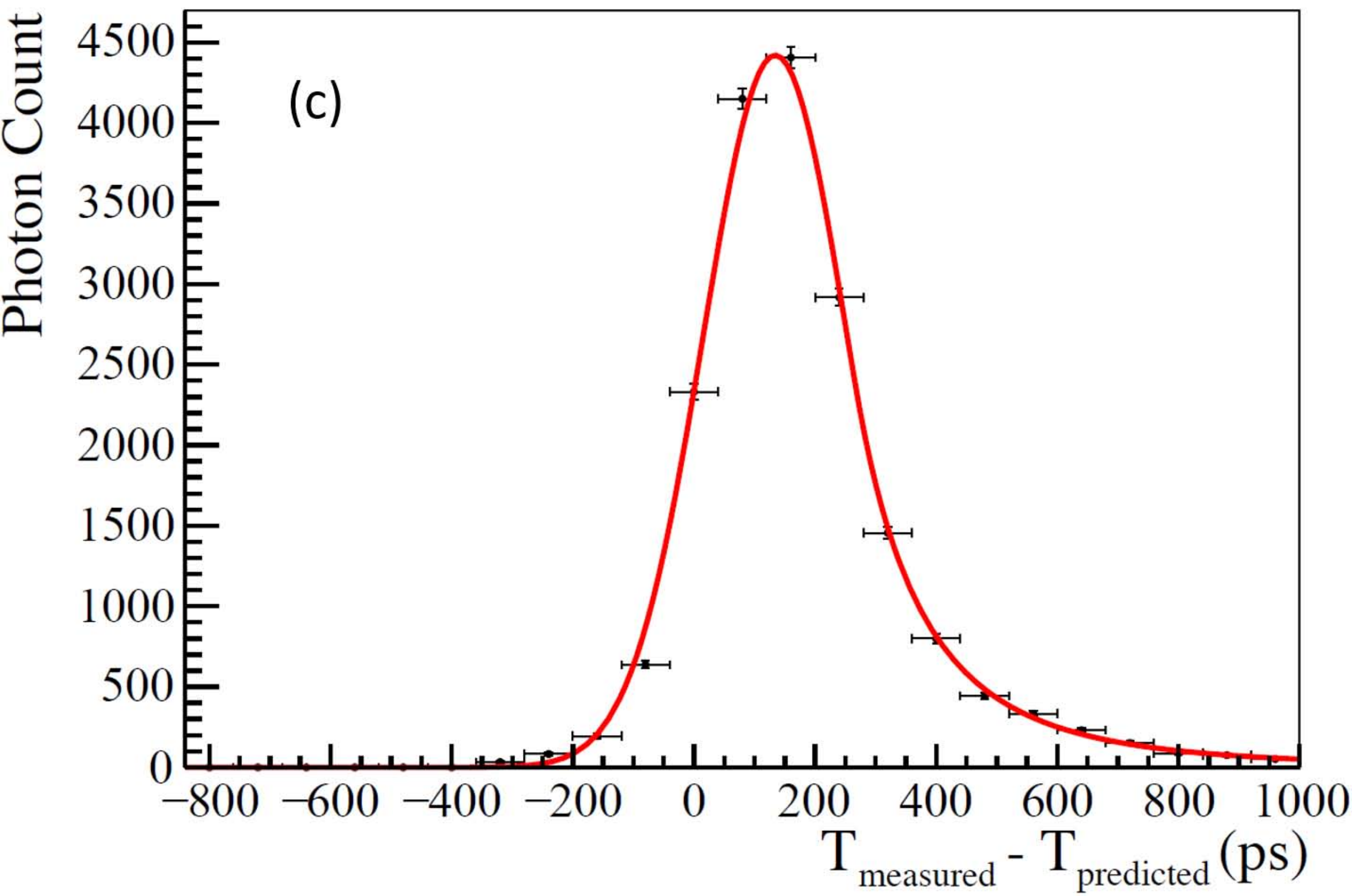} 
\caption{(a) The time-of-arrival of single Cherenkov photons in the small-scale TORCH demonstrator 
for 5\,GeV/$c$ pions,  
relative to the T2 beam time-reference station,
as a function of detected 64-wide row pixel number for column 5  of the 8 $\times$ 64 MCP-PMT configuration. 
The overlaid lines 
represent the simulated patterns.
(b) A schematic of the photon paths assigned in the reconstruction which  correspond to the 
overlaid lines in (a), labelled according to the number of side reflections the photon 
undergoes. Paths first reflecting off the  edge closest to the beam are shown by 
dashed lines, and their label has a prime. 
(c) An example plot of 
residuals  between the measured 
times of arrival from the predicted mean distribution for the 0 and 1$^\prime$ reflections.
}
\label{fig:timeprojection}
\end{figure}

For each single 4-wide or 8-wide pixel column, 
the measured MCP time-stamp for each cluster is plotted relative to the 
time-stamp of the downstream 
borosilicate station (T2) versus the measured 64-wide row position.  An example distribution 
for the  8 $\times$ 64 MCP-PMT configuration is 
shown in Fig.\,\ref{fig:timeprojection}\,(a) showing several orders of reflection, and 
demonstrating good agreement when compared to simulation.
 Figure \ref{fig:timeprojection}\,(b) highlights the corresponding photon paths.
The TORCH simulation uses GEANT4\,\cite{ref:GEANT4} to model the optical processes, including scattering, 
chromatic dispersion,  and the response of the MCP-PMT and electronics readout. 
Various sources of inefficiency are included, such as the surface roughness of the quartz and the 
MCP-PMT quantum and collection efficiencies.  
An example plot of 
residuals between the measured times of arrival from the mean for the primary reflections and
for a given MCP column   is shown in  Fig.\,\ref{fig:timeprojection}\,(c).
The tails are attributed to imperfect calibrations and photoelectron backscattering from the MCP top surface.
Core distributions have resolutions (sigmas) of approximately  100 - 125\,ps (which is photon energy and MCP-PMT column dependent).
The timing resolution of the timing reference is $\sim$43\,ps and, when subtracted in quadrature, gives  the 
 time resolutions presented in Table\,\ref{tab:Jun18Results}. These measurements   are
approaching the target resolution of 70\,ps per photon.
Future improvements will be made,  such as incorporating 
charge to width calibrations of the front-end electronics and reducing the current limitation imposed by the
100\,ps time binning of the HPTDC.

\begin{table}[htb]
\centering
\begin{tabular}{c|c|c}
MCP & $\sigma_{TORCH}$ & $\sigma_{TORCH}$ \\
Column & Pions$\,\rm{(ps)}$ & Protons$\,\rm{(ps)}$ \\
\hline
 &  & \\
1 &  {110.6} $\pm${1.2} &  {112.7} $\pm${1.4} \\
2 &  {101.7} $\pm${1.2} &  {110.6} $\pm${1.4} \\
3 &  {101.5} $\pm${1.2} &  {110.6} $\pm${1.4} \\
4 &  {105.5} $\pm${1.2} &  {106.2} $\pm${1.4} \\
5 &  {83.8} $\pm${1.3} &  {91.0} $\pm${1.4} \\
6 &  {101.3} $\pm${1.2} &  {103.4} $\pm${1.2} \\
7 &  {90.3} $\pm${1.2} &  {87.5} $\pm${1.4} \\
8 &  {112.4} $\pm${1.1} &  {102.8} $\pm${1.4}
\end{tabular}
\caption{The single-photon time resolutions measured for the $8 \times 64$ MCP-PMT
of the small-scale demonstrator.   
The MCP column numbers match those shown in Fig.\,\ref{fig:hits}(b). The 
quoted errors are purely statistical.}
\label{tab:Jun18Results}
\end{table}

\section{Beam tests with the large-scale TORCH demonstrator}

\begin{figure}
    \centering
\includegraphics[width=0.42\linewidth]{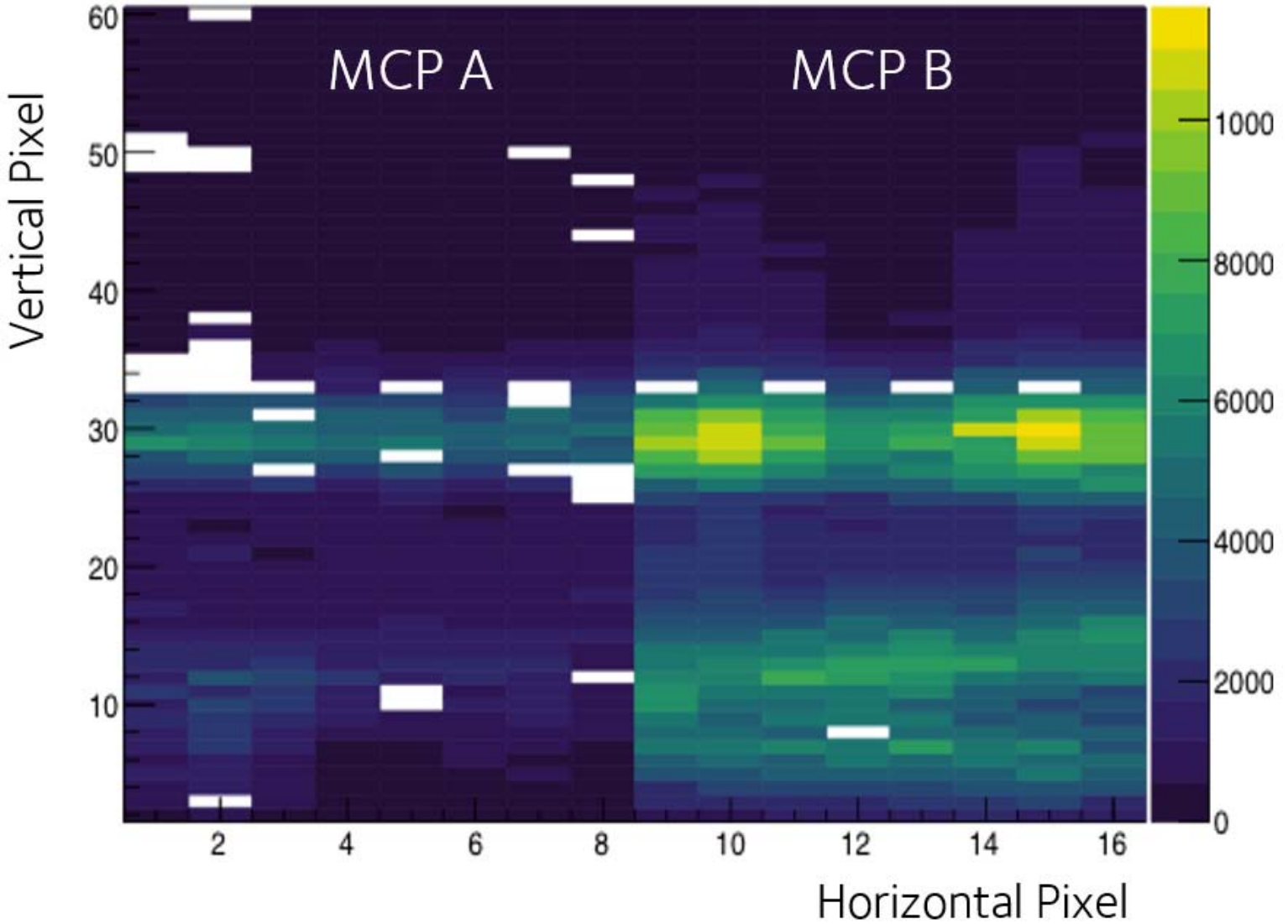}
    \includegraphics[width=0.42\linewidth]{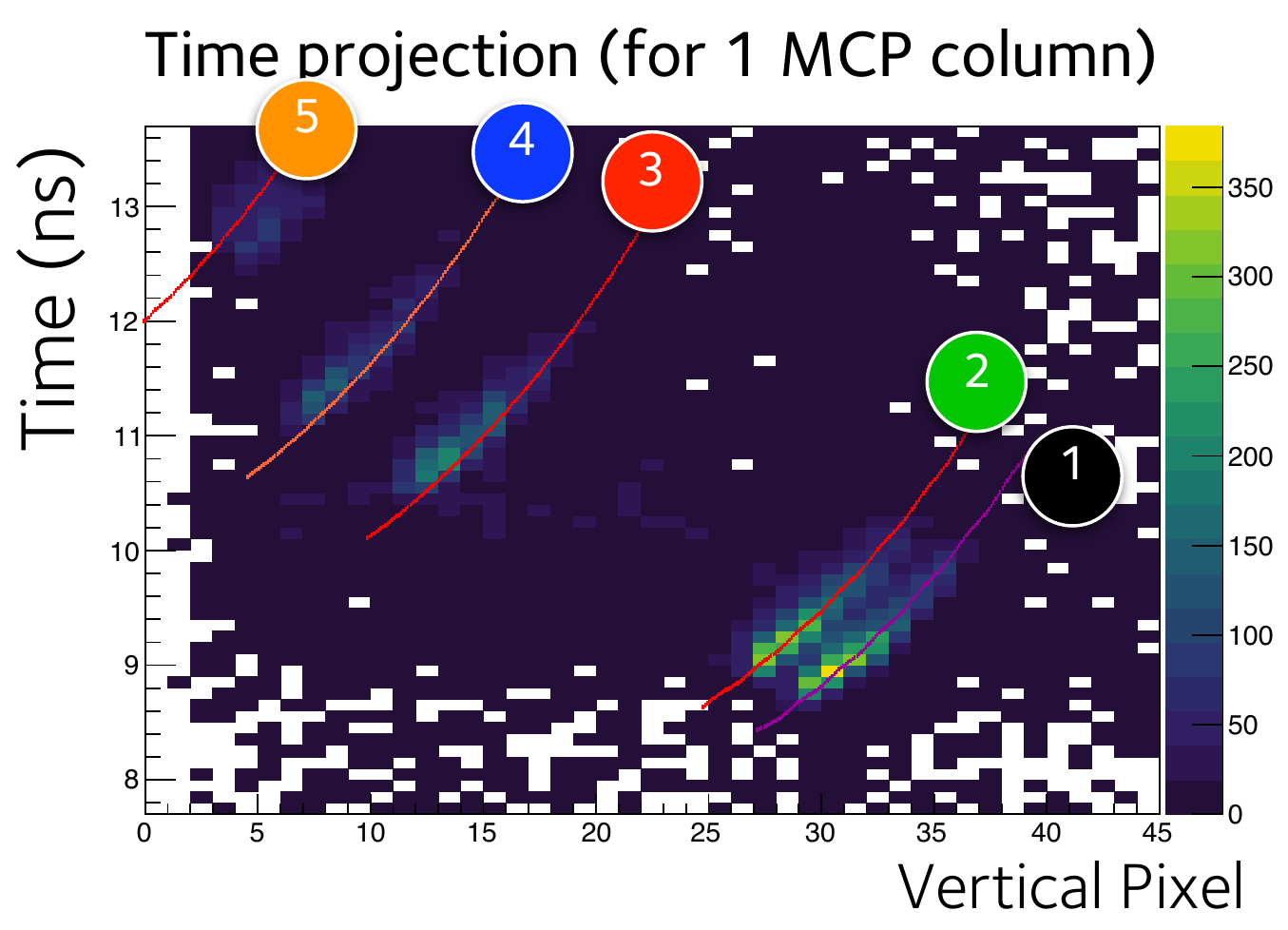}
    \includegraphics[width=0.14\linewidth]{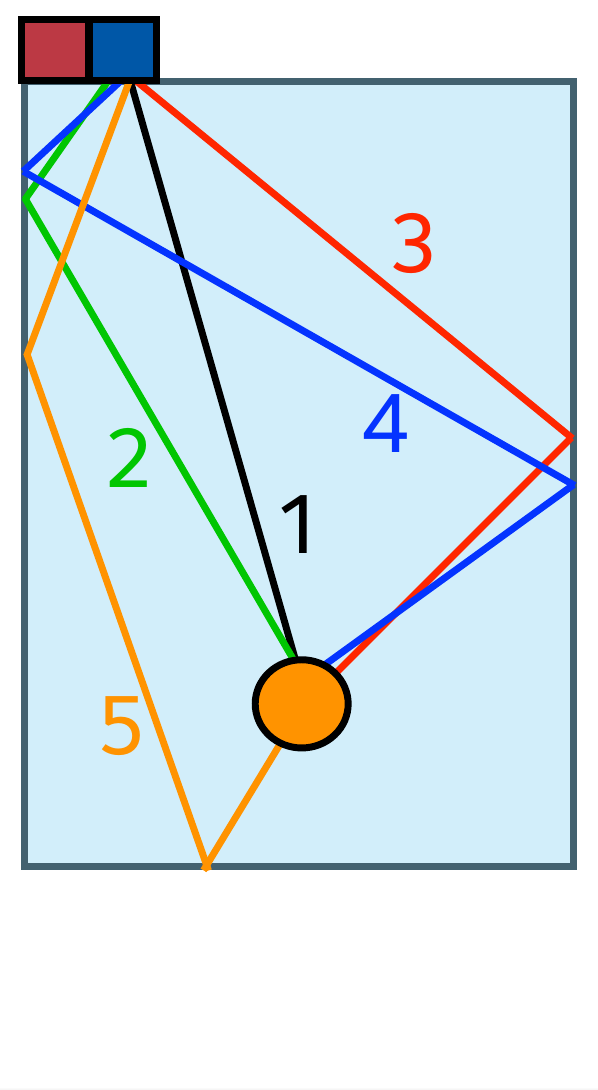}
    \caption{(Left) The spatial distribution of hits on the two MCP-PMTs of the half-length TORCH module,
labelled MCP-A and MCP-B. 
16 channels were read out for the time reference,
indicated by the central white band of empty hits (vertical pixel row 32).
(Middle) The time projection of a single MCP-PMT column
versus hit pixel number, with the photon trajectories 
labelled as shown in the diagram on the right.}
    \label{fig:hitmap}
\end{figure}

The half-length TORCH module was tested in a 8\,GeV/$c$ mixed pion-proton 
beam in October 2018\,\cite{ref:Hancock}. The beam infrastructure was
as described in Sec.\,\ref{sec:small-scale}.  
Two MCP-PMTs were instrumented and
the spatial distribution of measured hits is shown in Fig.~\ref{fig:hitmap}(a). 
The left MCP-PMT (labelled MCP-A)  has a  quantum efficiency which peaks at only $\sim$13\% 
and has a number of dead pixels due to disconnected 
NINO wire bonds. Hence, the analysis for the study presented below was based only on the
right MCP-PMT (labelled MCP-B), which has a quantum efficiency peaking at $\sim$18\%. 

An example time-projection plot for a single column of MCP-B  
as a function of pixel hit  is shown
in  Fig.~\ref{fig:hitmap}(b), along with the predictions from the reconstruction. 
It can be seen that several orders of side 
reflection are cleanly separated. As for the small-scale demonstrator,  
the widths of each order are measured from fits to determine the single-photon 
time resolution for that MCP column.
The  time resolution is then corrected for the uncertainty arising from the time reference pulse ($\sim43$\,ps) 
and from the beam spread uncertainty 
($\sim 15-30$\,ps, which is dependent on the distance between the beam entry point and the MCP-PMT plane). 

\begin{figure}
    \centering
    \includegraphics[width=0.42\linewidth]{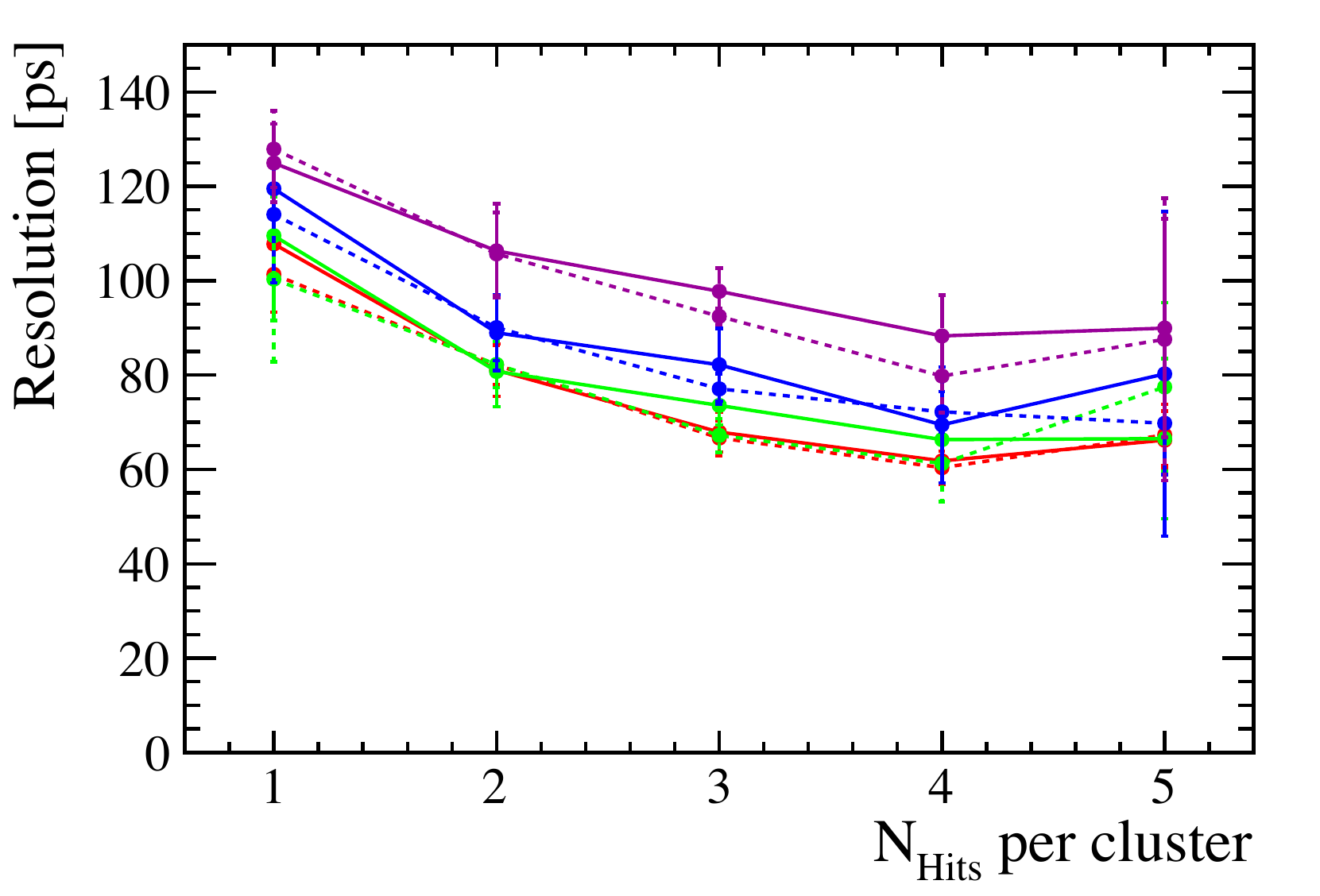}
    \includegraphics[width=0.14\linewidth]{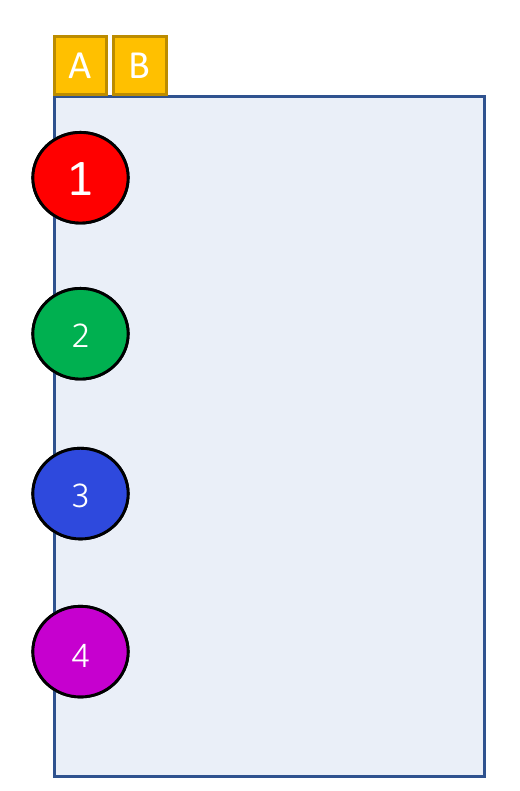}
    \caption{ The single-photon time resolution 
for the half-length TORCH module as a function of the number of hits per cluster,
as determined for the beam positions colour-coded in the diagram on the right. The solid lines are for 
pions, the dotted lines for protons. The vertical positions from the MCP-PMT plane
are 17.5\,cm (top), 45.3\,cm, 73.2\,cm and 101.0\,cm (bottom).}
    \label{fig:time_resolution}
\end{figure}

The vertical distance of the beam incidence position in the quartz
to the MCP-PMT plane was varied from 17.5\,cm to 101.0\,cm in four steps.
The measured time resolutions for first-order reflections 
are shown in Fig.~\ref{fig:time_resolution} as a function of the number of hits per cluster,
corresponding to all photon clusters that contain between 2 and 5 hits.
 Some degradation of the time resolution 
with distance of photon propagation is observed, however this reduces as the size of the cluster increases.  
The single-photon time resolution approaches or matches the design goal of 70\,ps for specific beam 
positions and cluster sizes.

\begin{figure}
    \centering
    \includegraphics[width=0.4\linewidth]{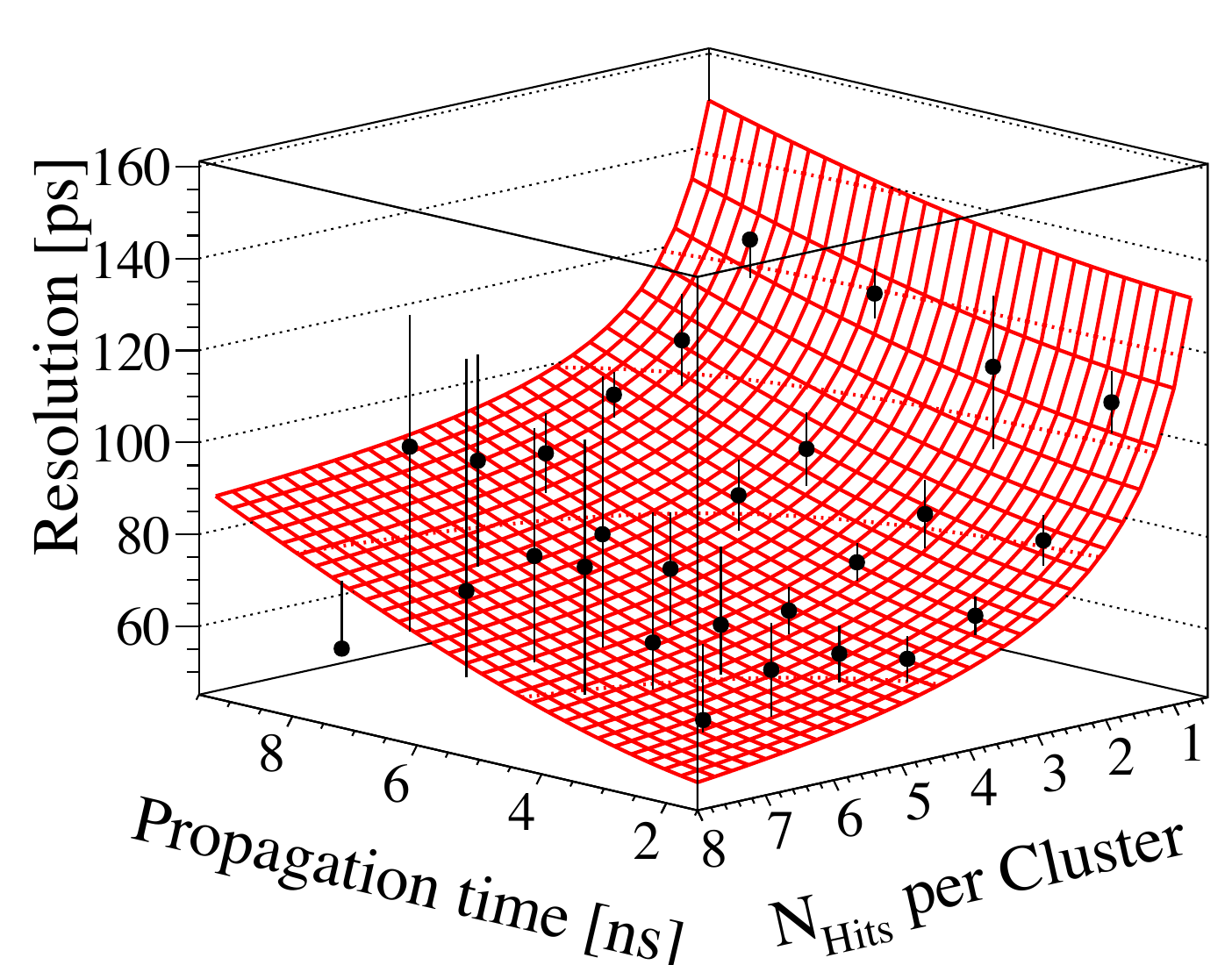}
    \caption{The TORCH time resolution   as a function of the photon propagation time and 
number of hits per cluster. The overlaid surface is the result of a 2D fit, described in the text.}
    \label{fig:time_resolution_2D}
\end{figure}

The  contributions to the timing resolution, $\sigma_{\text{TORCH}}$,  can be attributed
to the sources defined in the following expression: 
\begin{equation}
    \sigma_{\text{TORCH}}^{2} = \sigma^{2}_{\text{const}}  + \sigma^{2}_{\text{prop}}(t_{P})  + \sigma^{2}_{\text{RO}}(N_{\text{Hits}}). 
\end{equation}
Here the constant contribution, $\sigma_{\text{const}}$, includes sources such as the intrinsic 
MCP-PMT time resolution,
the term $\sigma_{\text{prop}}(t_{P})$ characterises the contributions that grow linearly
with photon propagation time, $t_{P}$, such as dispersive effects, 
and finally the term $\sigma_{\text{RO}}(N_{\text{Hits}})$ 
which is inversely proportional to the square root of the number of hits in a cluster, $N_{\text{Hits}}$. 
A two dimensional fit, shown in Fig.~\ref{fig:time_resolution_2D}, has been performed to the time resolution as a function of $t_{P}$ and $N_{\text{Hits}}$, with best-fit values as follows:
\begin{equation*}
    \sigma_{\text{const}} = 33.0\pm7.1\,ps,
\end{equation*}
\begin{equation*}
    \sigma_{\text{prop}}(t_{P}) = (7.8\pm0.7)\times10^{-3} \times t_{P}~[\text{in} \,ps], ~~\text{and}
\end{equation*}
\begin{equation*}
    \sigma_{\text{RO}}(N_{\text{Hits}}) = \frac{100.5\pm5.7}{\sqrt{N_{\text{Hits}}}}\,ps.
\end{equation*}
 Further studies of these contributions are currently underway, and
improvements can be expected with refined electronics calibrations in the future.

\begin{figure}
    \centering
    \includegraphics[width=0.8\linewidth]{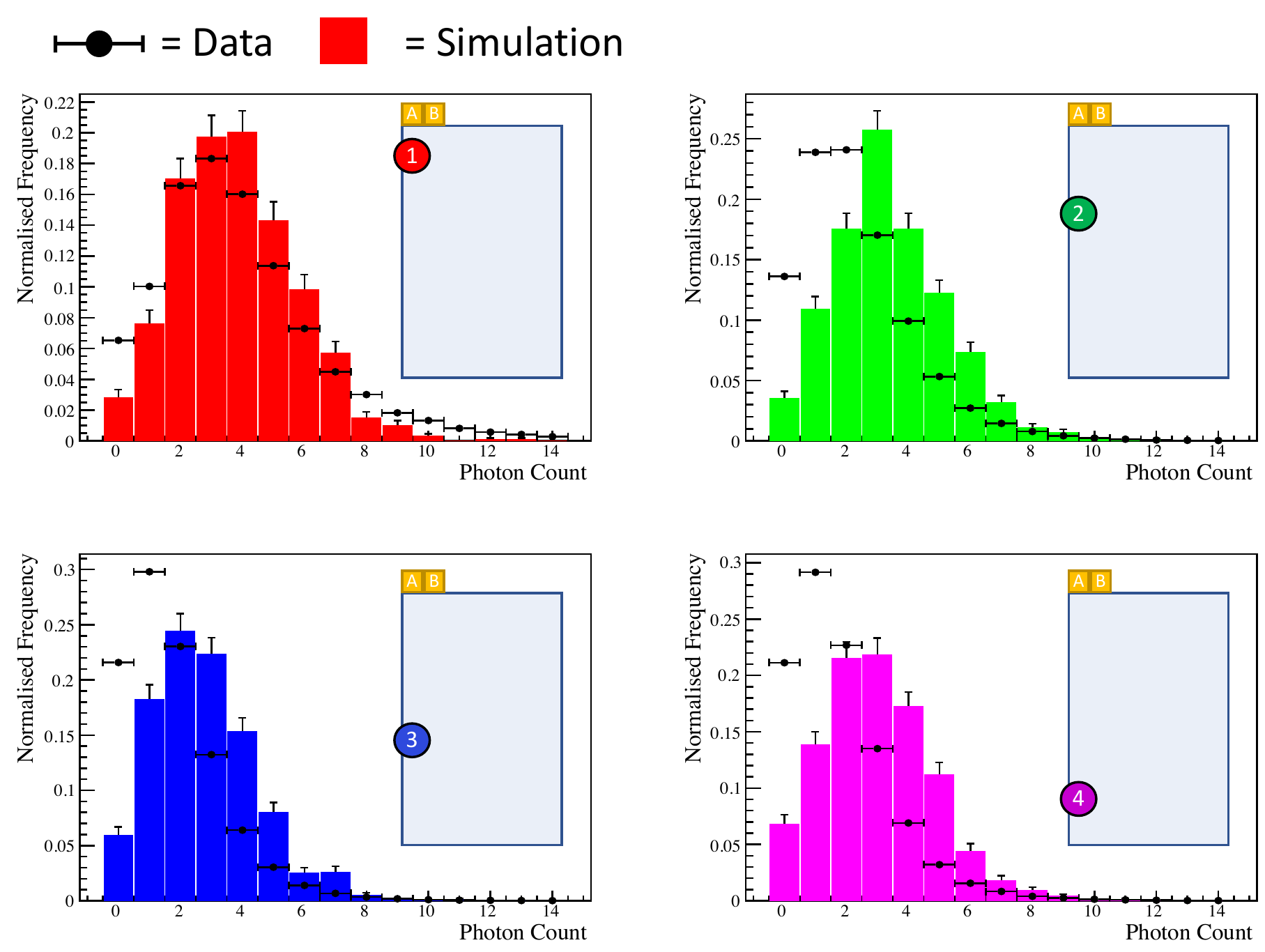}
    \caption{Photon yields in data (points) and simulation (histograms) 
when the beam is positioned 17.5\,cm (top left), 45.3\,cm (top right), 73.2\,cm (bottom left)
and 101.0\,cm (bottom right) from the MCP-PMT plane.}
    \label{fig:photon_counting}
\end{figure}

The number of photon  clusters measured in the half-length demonstrator are  compared to simulation
in Fig.~\ref{fig:photon_counting}. 
Reasonable agreement is seen when the beam is close to the MCP-PMTs, but discrepancies are observed 
when the photon path becomes longer,  an effect which is currently under study. 
As the half-length prototype is instrumented with only two MCP-PMTs out of the design total of eleven, the light yield is expected to improve by a factor of 5.5 in the fully instrumented module.
Additionally, the peak quantum efficiencies of the final MCP-PMTs are expected to have a factor $\sim$2 improvement, further increasing the photon yield to meet the 30 targeted.

\section{Simulation of TORCH at LHCb}

\begin{figure}
\centering
\includegraphics[width=0.75\linewidth]{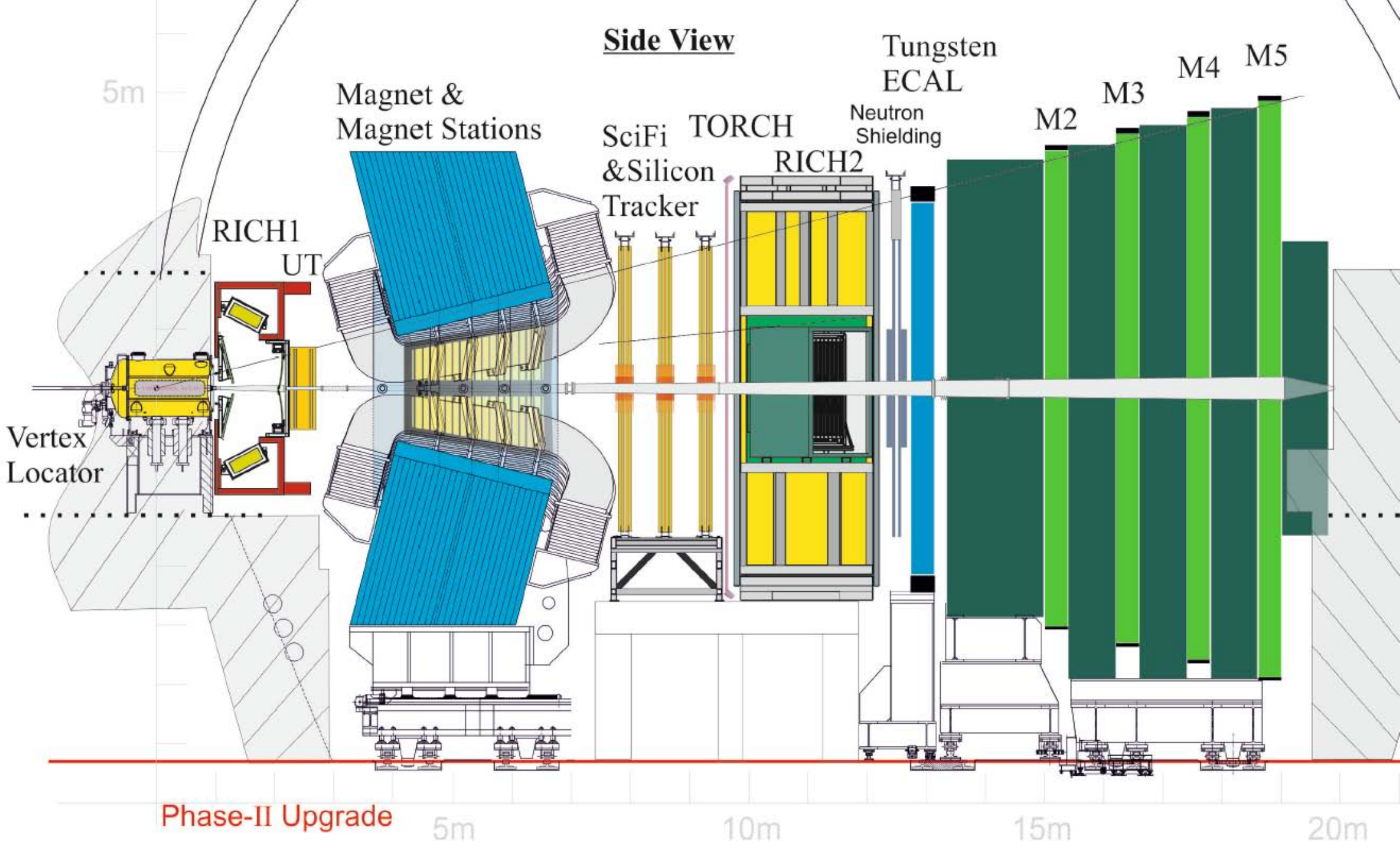}
\caption{A schematic of the LHCb experiment, showing TORCH located directly upstream of the
 RICH\,2 detector.}
\label{fig:LHCb-schematic}
\end{figure}

An exciting application of TORCH  is for the 
LHCb Upgrade II experiment, where the detector would occupy an area of
5 $\times$ 6 \,m$^2$ in front of the current RICH\,2 detector
 to complement the experiment's particle
identification (PID) capabilities, covering up to 10\,GeV/$c$ and beyond\,\cite{ref:upgradeTDR, ref:physicscase}.
The full TORCH detector would comprise
 18 modules with 198 MCP-PMTs, and studies are ongoing to simulate  the TORCH performance in GEANT4.
The  proposed experimental location 
together with the modular arrangement are shown in Figs.\,\ref{fig:LHCb-schematic}
and \ref{fig:design_TORCH_module}, respectively.

\begin{figure}[htbp]
\centering 
\includegraphics[width=1.0\columnwidth]{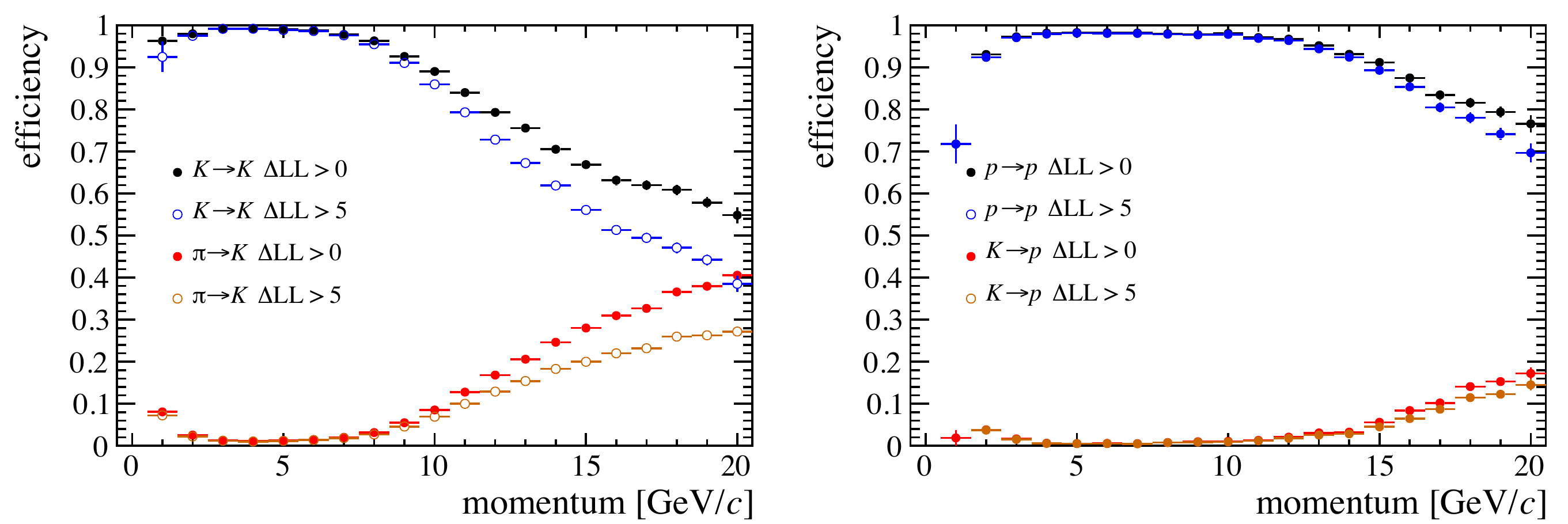}
\caption{\label{fig:LHCBefficiency} The efficiency in LHCb for TORCH (left) to positively identify 
kaons as a function of momentum and the probability that pions are misidentified as 
kaons and (right) the same plots for kaons and protons. 
The curves are for two different delta-log-likelihood cuts and for a luminosity of 2 $\times$ 10$^{33}$ cm$^{-2}$s$^{-1}$. }
\end{figure}

The TORCH PID performance has been determined for the LHCb Upgrade (Run 4) conditions at an
instantaneous luminosity of $\mathcal{L} = 2 \times 10^{33}\,\text{cm}^{-2}\,\text{s}^{-1}$. 
Figure\,\ref{fig:LHCBefficiency}  shows the efficiency for TORCH to positively identify 
kaons as a function of momentum and the probability that 
pions are misidentified as kaons. This demonstrates that the TORCH detector has excellent 
separation power in the
 range up to 10\,GeV/$c$ for kaons and pions, and up to 20\,GeV/$c$ for kaons and protons. 
This work will form the basis of a Technical Proposal to construct a 
full-scale TORCH detector for the start-up of LHC Run\,4, for installation in the Long Shutdown\,3 (LS3).

\section{Conclusions and future work}

TORCH is an innovative detector concept designed  to achieve a time-of-flight resolution of 15\,ps,
providing a $\pi -K$ separation up to 10\,GeV/$c$ momentum over a 10\,m flight path. 
A successful series of beam tests has been conducted with both a small-scale TORCH demonstrator and a half-length module.
Customised MCP-PMTs have been employed  which satisfy the TORCH requirements of lifetime, granularity and active area. 
Single-photon time resolutions have been measured in the range 60--130\,ps, 
achieving the goal of 70\,ps in some configurations. 
Further laboratory tests are ongoing to independently verify the contributions to the time resolution, and improved electronics 
calibrations are expected to further enhance the performance. 

A full-scale 18-module TORCH detector has been simulated in the LHCb 
experiment and these studies  indicate that significant improvements in the experiment's particle
identification capabilities can be expected. 
The half-length TORCH module will be instrumented with 10 MCP-PMTs, and will be 
 tested in the future months.

\acknowledgments

The support of the European Research Council through an Advanced Grant under the Seventh Framework Programme
(FP7), ERC-2011-AdG 299175-TORCH, and  the Science and Technology Research Council, UK, through grant number ST/P002692/ are acknowledged.
The Corresponding Author (NH)
would like to thank the organisers of  DIRC2019  for their support, and for hosting such an enjoyable  Workshop.

\end{document}